\documentclass{appolb}
\usepackage{epsfig}
\newcommand\beq{\begin{eqnarray}}
\newcommand\eeq{\end{eqnarray}}

\newcount\eLiNe\eLiNe=\inputlineno\advance\eLiNe by -1
\begin{document}
\title{PROBLEM OF NEUTRINO HANDEDNESS
IN WEAK INTERACTIONS}
\author{S. CIECHANOWICZ  \and W. SOBK\'OW
\address{Institute of Theoretical Physics, University of Wroc\l{}aw,
Pl. M. Born 9, PL-50-204~Wroc\l{}aw, Poland\\
e-mail: {\tt ciechano@rose.ift.uni.wroc.pl}  \\ e-mail: {\tt
sobkow@rose.ift.uni.wroc.pl}}
 \and  M. MISIASZEK
 \address{M. Smoluchowski
Institute of Physics, Jagiellonian University, ul. Reymonta 4,\\
PL-30-059 Krak\'ow, Poland \\
 e-mail: {\tt misiaszek@zefir.if.uj.edu.pl}}}
\maketitle
\begin{abstract}
In this paper we show that the neutrino observables,  including an
information about the transverse neutrino spin polarization, can
be sensitive to the effects coming from the interference terms
between the standard vector V, axial A couplings of L-handed
neutrinos and exotic scalar S coupling of
R-handed ones in the differential cross section. Our analysis is
based on  the electron neutrino-electron elastic scattering. This
reaction is considered at the level of the four-fermion point
interaction. Neutrinos are assumed to be massive and to be
polarized Dirac fermions coming from the Sun.
\end{abstract}


\section{Introduction}
According to the Standard Model (SM) of electro-weak interactions
\cite{Glashow,Wein,Salam} only vector V and axial A couplings of
Left-handed Dirac neutrinos are present in the weak interaction
processes. However on the other hand, a current precision of
measurements still allows a deviation from the standard V-A
structure of the charged and neutral weak interactions. There is
some space for the scenarios with the exotic scalar S, tensor T
and pseudoscalar  P couplings of the Right-handed Dirac neutrinos
beyond the  SM. The presence of new couplings would mean that the
charged and neutral weak interactions can be mediated by heavy,
charged, neutral bosons with spin-zero, one and two. \par The
current upper limits on  the all non-standard couplings, obtained
from the normal and inverse muon decay, are presented in the Table
\ref{table1} \cite{Data}. The transition amplitude for the muon
decay $ \mu^{-} \rightarrow \overline{\nu}_{e } + \nu_{\mu}+
e^{-}$ is follows: \beq M_{\mu^-} &=&
\frac{4G_{F}}{\sqrt{2}}\sum_{\gamma = S, V, T, }\sum_{\epsilon,
\mu = R, L}g_{\epsilon \mu}^{\gamma}\bigg<
\overline{e}_{\epsilon}|\Gamma^{\gamma}| (\nu_{e})_{n}\bigg>
\bigg<(\overline{\nu}_{\mu})_{m}|\Gamma_{\gamma}| \mu_{\mu}\bigg>.
\eeq Here, $\gamma= S, V, T$ indicates the type of weak
interaction, i.e.  scalar S, vector V, tensor T; $\epsilon, \mu=L,
R$ indicate the chirality of the electron or muon and the neutrino
chiralities $n, m$ are uniquely determined for given $\gamma,
\epsilon, \mu$. It means that the neutrino chirality $n$ or $m$ is
the same as the associated charged lepton  for the V interaction,
and opposite for the S, T interactions.
 In the SM, all of coupling
constants are set to zero by hand, except for $g_{LL}^{V}$ which
is set to one.
\begin{table}
\caption{\label{table1} Current limits on the non-standard
couplings}
\begin{center}
\begin{tabular}{|c|c|c|}
  \hline
  Coupling constants & SM & Current limits \\
  \hline
   $|g_{LL}^V|$ & $1$ &   $>0.960$ \\
    $ |g_{LR}^V|$ & $0$ & $<0.060$ \\
    $|g_{RL}^V|$ & $0$ & $<0.110$ \\
    $|g_{RR}^V|$ & $0$ & $<0.039$ \\
    \hline
  $|g_{LL}^{S}|$ & 0 & $<0.550$\\
  $|g_{LR}^{S}|$& 0 & $<0.125$\\
  $|g_{RL}^{S}|$& 0 & $<0.424$\\
  $|g_{RR}^{S}|$& 0 & $<0.066$\\
  \hline
   $|g_{LL}^{T}|$& 0 & $0$\\
   $|g_{LR}^{T}|$& 0 & $<0.036$\\
   $|g_{RL}^{T}|$& 0 & $<0.122$\\
   $|g_{RR}^{T}|$& 0 & $0$\\
  \hline
\end{tabular}
\end{center}
\end{table}
\par The high-precision measurements of various observables at low
energy area could observe or constrain physics beyond the SM. The
suitable quantities should consist only of the interference terms
between the standard $(V, A)_{L}$ and exotic $(S, T, P)_{R}$
couplings, which do not vanish in the massless neutrino limit. We
mean here the neutrino observables, including the information
about the transverse neutrino spin polarization (TNSP), both
T-even and T-odd components. These observables vanish in the SM,
so non-zero values would be a clear signature of the exotic
Right-handed weak interactions (ERWI).
 Currently  direct  tests are still difficult, but the process of the
neutrino-electron elastic scattering (NEES) could help in
searching for  new effects.
\par  In this paper we indicate the new tests of the Lorentz
structure  of the weak interactions using the NEES as a detection process.

\par
 We use the system of natural units with $\hbar=c=1$, Dirac-Pauli
representation of the $\gamma$-matrices and the $(+, -, -, -)$
metric \cite{Mandl}.

\section{ Helicity, chirality and exotic interactions}

As is known, measuring oscillation tells us the difference of
the neutrino mass square, so the number we find is a lower limit on the
neutrino mass squared. If the mass of the neutrino does not equal
zero, however small the mass is, the neutrino has two helicity
eigenstates. The helicity operator is not Lorentz invariant. When
taking a Lorentz boost with a speed faster than the neutrino, the
helicity of the neutrino would change its sign in the new
reference frame. The helicity of a neutrino depends on the
projection of its spin along its momentum. The massive neutrino
can be polarized perpendicular to its momentum, that is not
possible for massless particles. Even in the limit of massless
neutrino, a beam of neutrinos with non zero transverse
polarization is still a mixture of right- and left-handed
chirality states. A transversely polarized neutrino beam is not
chiral.
\par The chirality is a good quantum number only if the particle is massless.
This is because $\gamma_5$ does not commute with the Hamiltionian 
\beq i\hbar \frac{d}{dt} \gamma_5 = [ \gamma_5,  \, H] = 2 m c^2 \gamma_5 \beta, 
\eeq and hence the chirality is conserved only for a massless  fermion. 
\par The presence of the R-handed Dirac neutrinos allows to define
the Dirac mass term: \beq {\cal L_{D}} &=&
m_{D}(\overline{\Psi_{L}}\Psi_{R} + \overline{\Psi_{R}}\Psi_{L}).
\eeq This term couples left- and right-handed components of the
same field $\Psi$. It is the term that flips the chirality of a
particle. So, right-handed neutrino exists in a updated model, but
does not interact.
\par Assuming Lorentz invariance of the theory, one obtains the five bilinear Lorentz
covariants;  $\overline{u}u$ - the scalar $S$,
$\overline{u}\gamma^{5}u$ - pseudoscalar $P$,
$\overline{u}\gamma^{\mu}u$ - vector $V$,
$\overline{u}\gamma^{5}\gamma^{\mu}u$ - pseudovector $A$ and
$\overline{u}\sigma^{\mu \nu}u$ - tensor $T$ (antisymmetric tensor
of second rank).  These covariants may be expressed in terms of
the L- and R-handed chirality states. The $(V, A)$ couplings
conserve the initial particle chirality and $(S, T, P)$ couplings
flip chirality. For example, in the case of the $V$ and $S$
couplings for the neutrino-electron scattering ($\nu_{e } +
e^{-}\rightarrow \nu_{e } + e^{-}$), one gets, respectively: \beq
\overline{u}_{\nu_e'}\gamma^{\mu}P_{L}^{2}u_{\nu_e} & = & u_{\nu_e'}^{\dagger}P_{L}\gamma^{0}\gamma^{\mu}P_{L}u_{\nu_e},\\
\overline{u}_{\nu_e'}P_{L}^{2}u_{\nu_e} & = &
u_{\nu_e'}^{\dagger}P_{R}\gamma^{0}P_{L}u_{\nu_e},\nonumber \eeq where
\begin{equation} P_{L, R}= \frac{1}{2}(1\mp \gamma^{5}).
\end{equation}
 We see that the chiralities of the initial   and final
neutrino are identical (L-handed) in the standard $V$ weak
interaction, while the initial L-handed neutrino becomes the
outgoing R-handed one in the exotic $S$ weak interaction.
\par C.S. Wu pointed out that the exotic $(S,T,P)_{R}$ weak interactions
may be responsible for the negative electron helicity observed in
$\beta$-decay.
\par In muon decay experiments, we measure the energy spectrum of the
final state charged lepton \cite{Data}. The theoretical value of
the Michel $\rho$ parameter in terms of the all possible coupling
constants is \cite{loh95}
\begin{eqnarray}
\rho \, = \, \frac{3}{4} (|g^V_{LL}|^2 \, + \,
\frac{1}{4}|g^S_{LL}|^2)
+ \frac{3}{4} (|g^V_{RR}|^2 \, + \, \frac{1}{4}|g^S_{RR}|^2) \\
\nonumber + \frac{3}{16} (|g^S_{RL} - 2g^T_{RL}|^2 \, + \,
|g^S_{LR} - 2g^T_{LR}|^2).
\end{eqnarray}
We see that the measurement $\rho = 0.75$ in itself does not
determine the type of interaction. In the extreme, it is
compatible with a pure scalar type interaction $
g^S_{LL} = 2,g^V_{LL} = g^V_{RR} = g^S_{RR} = g^S_{RL} = g^S_{LR} = 
g^T_{RL} = g^T_{LR} = 0$. We can resolve this problem, however, by including
the data from the neutrino-electron scattering (i.e. the inverse
muon decay) \cite{fetscher86,charmII}.
\par
The Michel-Wightman density matrix \cite{Michel}
$\Lambda_{\nu}^{(s)}$ for the polarized massive neutrino, even for
infinitesimally small mass $m_{\nu}$ , remains finite including
the transverse component of neutrino polarization: \beq
\lim_{m_{\nu}\rightarrow 0} \Lambda_{\nu}^{(s)}
 & = & \mbox{} \left[1 + \gamma_{5}\left(\frac{\mbox{\boldmath
$\hat{\eta}_{\nu}$} \cdot{\bf q}}{|{\bf q}|} - (\mbox{\boldmath
$\hat{\eta}_{\nu}$} - \frac{(\mbox{\boldmath
$\hat{\eta}_{\nu}$}\cdot{\bf q}){\bf q}}{|{\bf q}|^{2}})\cdot
\mbox{\boldmath
$\gamma$}\right)\right]\left(q^{\mu}\gamma_{\mu}\right), \eeq
where $\mbox{\boldmath $\hat{\eta}_{\nu}$}$ - the unit vector of
the
 initial neutrino polarization in its rest frame, ${\bf q}$ - the momentum of initial
 neutrino.

\section{Electron neutrino-electron scattering}

\begin{figure}
\begin{center}
\includegraphics[width=11cm,angle=0]{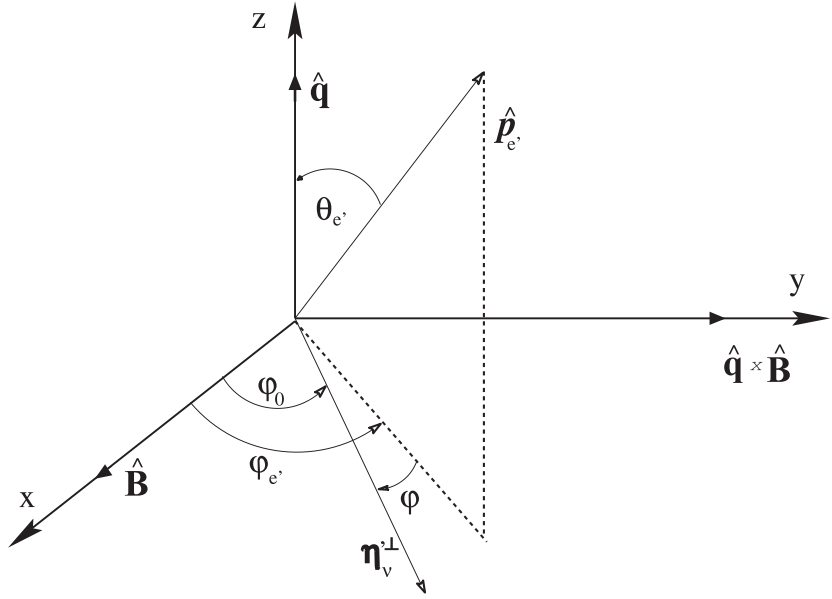}
\end{center}
\caption{The production plane of $\nu_e$ neutrinos,  reaction
plane for the $\nu_e e^-$ scattering with the transverse neutrino
polarization vector $\mbox{\boldmath $\eta_{\nu}^{' \perp}$}$.
$\theta_{e'}$ is the angle between the direction of the outgoing
electron momentum $ \hat{\bf p}_{e'}$ and the direction of the
incoming neutrino momentum $\hat{\bf q}$ (recoil electron
scattering angle). $\phi_{e'}$ is the angle between the production
plane and the reaction plane. $\phi$ is the angle between the
reaction plane and the transverse neutrino polarization vector and
is connected with the $\phi_{e'}$ in the following way;
$\phi=\phi_{0}-\phi_{e'}$. $\phi_{0}$ is the angle between the
production plane and the transverse neutrino polarization vector.}
 \label{wsp}
\end{figure}

In our considerations, we analyze the polarized electron neutrino
beam coming from the Sun.
The incoming electron neutrino flux is  the mixture of the
L-handed neutrinos produced in the standard $V-A $ charged weak
interaction and the R-handed ones produced by the spin flip
mechanism.
This mixture is detected in the neutral (NC) and charged current
(CC) weak interaction. One assumes that the incoming L-handed
neutrinos are detected in the standard V-A  weak interaction,
while the initial R-handed neutrinos are detected in the exotic S
one. In the final state all the neutrinos are L-handed. We analyze
a minimal version of the SM extension, because the admittance of
all exotic weak interactions does not change the qualitatively the
conclusions. The reaction plane is spanned by the direction of the
outgoing electron momentum $ \hat{\bf p}_{e'}$  and  of the
incoming neutrino momentum $\hat{\bf q}$, Fig.\ref{wsp}.
The transition amplitude for the $\nu_{e } e^{-}$ scattering is
follows: \beq M_{\nu_{e} e} &=&
\frac{G_{F}}{\sqrt{2}}\{(\overline{u}_{e'}\gamma^{\alpha}(c_{V}^{L}
- c_{A}^{L}\gamma_{5})u_{e}) (\overline{u}_{\nu_{e'}}
\gamma_{\alpha}(1 - \gamma_{5})u_{\nu_{e}})\nonumber \\ &  &
\mbox{} +
\frac{1}{2}c_{S}^{R}(\overline{u}_{e'}u_{e})(\overline{u}_{\nu_{e'}}
(1 + \gamma_{5})u_{\nu_{e}})\},
 \eeq
 where $ u_{e}$ and  $\overline{u}_{e'}$
$(u_{\nu_{\mu}}\;$ and $\; \overline{u}_{\nu_{\mu'}})$ are the
Dirac bispinors of the initial and final electron (neutrino)
respectively. $G_{F}= 1.16639(1)\times 10^{-5}GeV^{-2}$
\cite{Data} is the Fermi constant. The coupling constants are
denoted  as $c_{V}^{L} $, $c_{A}^{L}$ and $c_{S}^{R}$ respectively
to the incoming neutrino of L- and R-chirality.
\par The laboratory differential cross
section for the $\nu_{e}e^{-}$ scattering, in the limit of
vanishing neutrino mass, has  the form:
\beq \label{przekr} \frac{d^{2} \sigma}{d y d \phi_{e'}} & = &
\bigg(\frac{d^{2} \sigma}{d y d \phi_{e'}}\bigg)_{(V, A) } +
\bigg(\frac{d^{2} \sigma}{d y d \phi_{e'}}\bigg)_{(S)} +
\bigg(\frac{d^{2} \sigma}{d y d \phi_{e'}}\bigg)_{(V S)},
\\
\bigg(\frac{d^{2} \sigma}{d y d \phi_{e'}}\bigg)_{(V, A)} &=& B
\bigg\{ (1-\mbox{\boldmath $\hat{\eta}_{\nu}$}\cdot\hat{\bf q}
)\bigg[(c_{V}^{L} + c_{A}^{L})^{2} + (c_{V}^{L}-
c_{A}^{L})^{2}(1-y)^{2} \\
& & \mbox{} - \frac{m_{e}y}{E_{\nu}}((c_{V}^{L})^{2} -
(c_{A}^{L})^{2})\bigg]\bigg\},
\nonumber\\
\bigg(\frac{d^{2} \sigma}{d y d \phi_{e'}}\bigg)_{(S)} &=& \mbox{}
B \bigg\{(1+\mbox{\boldmath $\hat{\eta}_{\nu}$}\cdot\hat{\bf
q})\frac{1}{8}y(y+2\frac{m_{e}}{E_{\nu}})
 |c_{S}^{R}|^{2}\bigg\}, \\
\label{vlsr} \bigg(\frac{d^{2} \sigma}{d y d \phi_{e'}}\bigg)_{(V
S)} &=& \mbox{} B
\bigg\{\sqrt{y(y+2\frac{m_{e}}{E_{\nu}})}[-\mbox{\boldmath
$\hat{\eta}_{\nu}$}\cdot({\bf \hat{p}_{e'} \times
\hat{q}})Im(c_{V}^{L}c_{S}^{R*}) \\
& & \mbox{} + (\mbox{\boldmath $\hat{\eta}_{\nu}$}\cdot {\bf
\hat{p}_{e'}}) Re(c_{V}^{L}c_{S}^{R*})]  -
y(1+\frac{m_{e}}{E_{\nu}}) (\mbox{\boldmath
$\hat{\eta}_{\nu}$}\cdot\hat{\bf q})
Re(c_{V}^{L}c_{S}^{R*})\bigg\},\nonumber\\
  B & \equiv & \frac{E_{\nu}m_{e}}{(2\pi)^{2}}
\frac{G_{F}^{2}}{2}, \\ y & \equiv &
\frac{T_{e'}}{E_{\nu}}=\frac{m_{e}}{E_{\nu}}\frac{2cos^{2}\theta_{e'}}
{(1+\frac{m_{e}}{E_{\nu}})^{2}-cos^{2}\theta_{e'}}, \eeq where $y$
- the ratio of the kinetic energy of the recoil electron $T_{e'}$
to the incoming neutrino energy $E_{\nu}$ (the inelasticity parameter). 
It varies from $0 $ to
$2/(2+m_e/E_\nu) $. $m_{e}$ - the electron mass, $ \mbox{\boldmath
$\hat{\eta}_{\nu}$}\cdot\hat{\bf q}$ - the longitudinal
polarization of the incoming neutrino. \par  The
interference term, Eq. (\ref{vlsr}), includes only the transverse
components of the initial neutrino spin polarization, both
$T$-even and $T$-odd: \beq \label{inter} \bigg(\frac{d^{2}
\sigma}{d y d \phi_{e'}}\bigg)_{(VS)} &=&
B\bigg\{\sqrt{\frac{m_{e}}{E_{\nu}}y[2-(2+\frac{m_{e}}{E_{\nu}})y]}\\
&& \cdot |c_{V}^{L}||c_{S}^{R}||\mbox{\boldmath
$\eta_{\nu}^{'\perp }$}|cos(\phi-\alpha)\bigg\}, \nonumber \eeq
where $\alpha \equiv\alpha_{V}^{L} - \alpha_{S}^{R} $ - the
relative phase between the  $c_{V}^{L}$ and  $c_{S}^{R}$
couplings. This interference would be responsible for the
appearance of the azimuthal asymmetry of the electrons recoiled
after the subsequent neutrino scattering.  We see that the
interference contribution between the $c_{V}^{L}$ and $c_{S}^{R}$
couplings will be substantial at lower neutrino energies
$E_{\nu}\leq m_{e}$ but negligibly small at large energies and
vanishes for $\theta_{e'}=0$ or $\theta_{e'}=\pi/2$.
\par After integration over the $\phi_{e'}$, the interference term
vanishes and the cross section consists  only of two terms:
\beq \label{wyk2} \frac{d \sigma}{d y} & = & \bigg(\frac{d
\sigma}{d y}\bigg)_{(V, A)} + \bigg(\frac{d \sigma}{d y}\bigg)_{(S)},\\
\bigg(\frac{d \sigma}{d y}\bigg)_{(V, A)} & = & B' \bigg\{
(1-\mbox{\boldmath $\hat{\eta}_{\nu}$}\cdot\hat{\bf q})
\bigg[(c_{V}^{L} + c_{A}^{L})^{2}   + (c_{V}^{L}-
c_{A}^{L})^{2}(1-y)^{2} \nonumber
\\& & \mbox{} -  \frac{m_{e}y}{E_{\nu}}((c_{V}^{L})^{2} - (c_{A}^{L})^{2})\bigg]\bigg\}, \\
\bigg(\frac{d \sigma}{d y}\bigg)_{(S)} &=& \mbox{} B'
\bigg\{(1+\mbox{\boldmath $\hat{\eta}_{\nu}$}\cdot\hat{\bf
q})\frac{1}{8}y(y+2\frac{m_{e}}{E_{\nu}}) |c_{S}^{R}|^{2}\bigg\},
\eeq
where $B' = 2\pi B$.
\par If one assumes that  only L-handed neutrinos are produced in the standard
 $(V-A)$ and  non-standard $S $ weak interactions (pure L-handed neutrino beam),
 there is no interference
between the  $c_{V, A}^{L}$ and $c_{S}^{L} $ couplings in the
differential cross section, when $m_{\nu}\rightarrow 0$, and the
angular distribution of the recoil electrons has the azimuthal
symmetry.  We do not consider this scenario.

\subsection{Astrophysical consequences of R-handed neutrinos}
\par If a neutrino has a non-zero magnetic moment, the neutrino helicity
can be flipped when it passes through a region with magnetic field
perpendicular to the direction of propagation. It means that the
Left-handed neutrino that is active in SM would change into a
right-handed one $(\mbox{\boldmath
$\hat{\eta}_{\nu}$}\cdot\hat{\bf q}=1)$ that is sterile in SM:
\begin{eqnarray}
(\frac{d^{2}\sigma}{dyd\phi_{e'}})_{(V,A)}=\,(1-\mbox{\boldmath
$\hat{\eta}_{\nu}$}\cdot\hat{\bf q})\,\cdot\, f(E_{\nu}, y) &
=\,0.\label{eq:VA}
\end{eqnarray}
Solutions based on neutrino {}``spin flip'' in the Sun's magnetic fields
are proposed to explain\ the observed  solar neutrino deficit
\cite{key-1}. They are important alternatives to LMA MSW solution, 
because no one signature of the LMA has been observed. Dependence of the 
survival probability on energy and significant regeneration effect 
(day/night asymmetry) are not observed in solar neutrino detectors.
\par The scattering due to the photon exchange between a neutrino and a
charged particle in plasma leads to neutrino spin flip. 
The energy released in supernova implosion is taken partly 
away by sterile
neutrinos without further interactions. In this scenario the
neutrino magnetic moment should be bounded because of the observed
neutrino signal of SN 1987A\cite{key-2}. 
\par
\begin{figure}
\center
\includegraphics[angle=0,scale=.4]{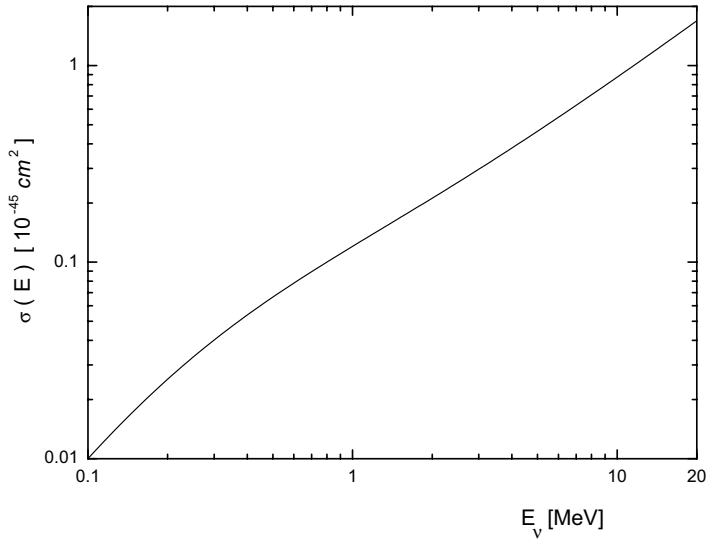}
\caption{Plot of the total cross section $\sigma(E)$ as a function
of Right-handed $(\mbox{\boldmath $\hat{\eta}_{\nu}$}\cdot\hat{\bf
q}=1)$ neutrino energy $E_\nu$ for the $(\nu_e e^{-})$
scattering. The exotic scalar coupling $c_S^R = g_{LL}^S + g_{LR}^S = 0.550
+0.125=0.675$ of right-handed electron neutrinos is used.} \label{Wy5MM}
\end{figure}

 Our paper shows that the participation of the exotic couplings 
of the right-handed
neutrinos can modify the both astrophysical considerations. The
right-handed neutrino is no longer {}``sterile''. The total cross
section for $\nu_{e}e^{-}$ scattering with the coupling constants
from the current data  can be calculated from our general formulas
(see Fig. \ref{Wy5MM}). In this scenario the right-handed
neutrinos can be detected by neutrino detectors and could help
simultaneously to transfer the energy to presupernova envelope.
Because of the SNO results \cite{sno} and comparison of signals in 
the Homestake \cite{sm5} and SuperKamiokande \cite{sm6} experiments,
we came to conclusion that solar neutrinos undergo the chirality conversion
and right-handed neutrinos do interact via the exotic (S,T,P) charged and 
neutral currents. 

\par If the conversions $\nu_{eL}\rightarrow \nu_{eR}$ in the Sun
are possible, the azimuthal asymmetry of the recoil electrons
generated by the interference terms between the standard $(V,
A)_{L}$ and exotic $S_{R}$ couplings should occur. If one assumes
that a survival probability
 for the left-handed $^{7}Be$-neutrinos  is equal to
 $P_{eL}=0.5$,
the value of the transverse neutrino polarization as a function of
this $P_{eL}$ will be large, $|\mbox{\boldmath $\eta_{\nu}^{'
\perp}$}|=2\sqrt{P_{eL}(1-P_{eL})}=1$. It means that in this case
$\mbox{\boldmath $\hat{\eta}_{\nu}$}\cdot\hat{\bf q}= 1-2\cdot
P_{eL}=0$, see Eq. (9) in \cite{Barbieri}. The equation on the
$|\mbox{\boldmath $\eta_{\nu}^{' \perp}$}|$ arises from the
density matrix for the relativistic neutrino chirality.

\section{Conclusions}

In this paper, we show that the scattering of the polarized
electron-neutrino beam on the unpolarized  electron target may be
sensitive to the interference effects between the L- and R-handed
neutrinos in the differential cross section for the
$(\nu_{e}e^{-})$ scattering. The terms with the interference
between the $(V, A)_{L}$ and $S_{R}$ couplings do not vanish in
the massless neutrino limit and depend on the azimuthal angle
between  the outgoing electron momentum and  the transverse
neutrino polarization. The non-vanishing interferences would
generate the azimuthal asymmetry of the recoil electrons. That
would be the direct signature of the R-handed neutrinos in the
$(\nu_{e} e^-)$ scattering. We also show that the participation
of the exotic couplings of the R-handed neutrinos can modify
astrophysical considerations. The R-handed neutrino is no longer
sterile.



This work was supported in part by the grant 2P03B 15522 of The Polish State 
Committee for Scientific Research and by the Foundation for Polish Science.

\end{document}